%% file: plb.tex
\documentclass[aps,prl,twocolumn,showpacs,groupedaddress]{revtex4}  % for review and submission
\usepackage{graphicx}  % needed for figures
\usepackage{dcolumn}   % needed for some tables
\usepackage{bm}        % for math
\usepackage{amssymb}   % for math

\begin{document}

% the following line is for submission, including submission to the arXiv!!
\hspace{5.2in} \mbox{Fermilab-Pub-10/038-E}

% remove the space for publication
%\vspace*{1.5cm}
\title{\boldmath Measurement of the dijet invariant mass cross section in $p\overline{p}$ collisions at $\sqrt{s} =$ 1.96 TeV \unboldmath
%\\ \vspace*{2.0cm}
}
\input list_of_authors_r2.tex  % input Dzero author list

%\author{Mandy Rominsky, Mike Strauss, Don Lincoln, Markus Wobisch \vspace*{0.5cm}}
% Also, during the review process, including collaboration review,
% include the header:
%\noindent Version: 3.7 \hfill Send comments to d0-run2eb-002@fnal.gov \\
%\noindent Version: 2.1 \\
%Author(s): Mandy Rominsky, Mike Strauss, Don Lincoln, Markus Wobisch by January 14, 2010
%\author{The D\O \ Collaboration\\ URL: {\it http://www-d0.fnal.gov}}
\date{February 24, 2010}

%\setpagewiselinenumbers
%\modulolinenumbers[3]
%\linenumbers
\begin{abstract}
%\vspace*{3.0cm}
The inclusive dijet production double differential cross
section as a function of the dijet invariant mass and
of the largest absolute rapidity of the two jets with the largest 
transverse momentum in
an event is measured in $p\overline{p}$ collisions
at $\sqrt{\rm s} = 1.96$ TeV using 0.7 fb$^{-1}$ of integrated 
luminosity collected with the D0 detector at the
Fermilab Tevatron Collider.  The measurement is
performed in six rapidity regions up to a maximum 
rapidity of 2.4.
Next-to-leading order perturbative QCD predictions are
found to be in agreement with the data.
\end{abstract}

\pacs{13.87.Ce, 12.38.Qk}
%\ http://www.aip.org/pacs/
\maketitle

%\section{\label{sec:intro}Introduction}

The dominant process contributing to the total inelastic cross section in
$p\overline{p}$ collisions at $\sqrt{s} = 1.96$ TeV is the
production of hadronic jets.  A measurement
of the dijet production cross section
as a function of the dijet invariant mass ($M_{\rm JJ}$) can
be used to test the predictions of perturbative
quantum chromodynamics (QCD), to constrain parton distribution
functions (PDFs) of the proton, and to look for signatures
of physics not predicted by the standard model.  This type of measurement
is sensitive to
quark compositeness, to extra spatial dimensions, and to undiscovered
heavy particles that decay into two
quarks \cite{composite,composite2,composite3,composite4,composite5,composite6,composite7,composite8}.
The distribution presented in this paper is particularly sensitive to the PDF of gluons at high proton momentum fraction, a region in which the gluon distribution is
weakly constrained.
Previous measurements of the dijet invariant mass dependent 
cross section in this energy
regime restricted the rapidity of
the jets to $|y| < 1.0$ \cite{runIpaper,dmr1,cdfrun2}
where $y=0.5\ln{[(E+p_L)/(E-p_L)]}$, $E$ is the energy of the jet, and
$p_L$ is the component of momentum along the direction of the proton beam.

In this Letter, we present a measurement of the double differential
dijet production cross section as a 
function of the dijet invariant mass and the
variable $|y|_{\rm max}$, for $0< |y|_{\rm max} < 2.4$.
The dijet invariant mass is computed from the four momenta
of the two jets with largest transverse momentum ($p_T$) with respect
to the beam direction. Both jets are required to have
$p_T >40$ GeV.  The variable $|y|_{\rm max}$ is defined as
$|y|_{\rm max} = {\rm max}(|y_1|,|y_2|)$ where
$y_1$ and $y_2$ are the rapidities of the two jets with the
largest $p_T$. The cross section results are corrected for
instrumental effects and presented at the particle level, which
includes energy from stable particles, the underlying event,
muons, and neutrinos, as defined in Ref.~\cite{Buttar:2008jx}.

%\subsection{\label{sec:level2}Second-level heading: Formatting}
% subsections are not used for PRL papers
%This file may be formatted in both the \texttt{preprint} and
%\texttt{twocolumn} styles. \texttt{twocolumn} format may be used to
%mimic final journal output. Either format may be used for submission
%purposes; however, for peer review and production, APS will format the
%article using the \texttt{preprint} class option. Hence, it is
%essential that authors check that their manuscripts format acceptably
%under \texttt{preprint}. Manuscripts submitted to APS that do not
%format correctly under the \texttt{preprint} option may be delayed in
%both the editorial and production processes.
%\section{\label{sec:anal}Analysis}

This measurement uses approximately
0.7 fb$^{-1}$ of integrated luminosity collected
with the D0 detector \cite{d0det} at the Fermilab Tevatron Collider in
$p\overline{p}$ collisions at $\sqrt{s} = 1.96$ TeV during 2004--2005.
Outgoing partons created in the scattering process hadronize to
produce jets of particles that are detected in the finely
segmented liquid-argon and uranium calorimeters which cover most of
the solid angle.  The central calorimeter (CC) covers the
pseudorapidity region $|\eta|$ up to 1.1 ($\eta = - \ln [ \tan ( \theta /2 )]$
where $\theta$ is the angle with respect to the proton beam direction) and the
two end calorimeters (EC) extend the coverage up to $|\eta| <  4.2$.
The intercryostat region (ICR) between the CC
and EC contains scintillator-based detectors to improve the energy sampling in this region.
Jets are reconstructed by clustering
energy deposited in the calorimeter towers using an iterative
seed-based cone jet algorithm including midpoints~\cite{d0jets}
with cone radius
$\mathcal{R} = \sqrt{(\Delta y)^2 + (\Delta \phi)^2} = 0.7$, where $\phi$ is
the azimuthal angle.  The $p_T$ of each jet is calculated using only
calorimeter information and the location of the $p\overline{p}$
collision.
The  measurement is performed in six rapidity
regions: $0 < |y|_{\rm max} \leq 0.4$,
$0.4 < |y|_{\rm max} \leq 0.8$, $0.8 < |y|_{\rm max} \leq 1.2$,
$1.2 < |y|_{\rm max} \leq 1.6$, $1.6 < |y|_{\rm max} \leq 2.0$,
and $2.0 < |y|_{\rm max} \leq 2.4$.

Events are required to satisfy jet $p_T$ or dijet invariant mass
dependent trigger requirements with minimum
dijet invariant mass thresholds.
Trigger efficiencies are studied by comparing observables in data
sets collected with higher trigger thresholds to those collected using
lower trigger thresholds
in regions where the lower threshold
trigger is 100\% efficient.  The trigger with
the lowest threshold is determined to be 100\% efficient
in the region of interest by comparing
it with sample of independently triggered muon events.
For $|y|_{\rm max}\leq1.6$, single jet triggers are used, while dijet invariant
mass triggers are used for $|y|_{\rm max} > 1.6$.
The measurement is only done in the kinematic regions where the trigger efficiency is $> 99\%$.

Events must satisfy data and jet quality requirements.
The position of the $p\overline{p}$ interaction is reconstructed
using a tracking system consisting of silicon microstrip detectors and
scintillating fibers located inside a solenoidal magnetic
field of approximately 2 T.
The position of this primary vertex along the beam line is required
to be within 50 cm of the detector center.  This requirement is $\approx 93\%$ efficient.
Requirements based on calorimeter shower shapes
are used to remove the remaining background due to electrons, photons, and
detector noise that mimic jets. The sample selection efficiency is
$>$ 99\% ($>$ 97.5\% for $0.8 < |y|_{\rm max} < 1.6$).
In order to suppress
cosmic ray events, the requirements $\not{\hspace{-.065in}}E_T/p_T^{\rm max} < 0.7$ for $p_T < 100$ GeV of the highest $p_T$ jet 
and $\not{\hspace{-.065in}}E_T/p_T^{\rm max} < 0.5$ otherwise
are applied, where $\not{\hspace{-.05in}}E_T$ is
the transverse component of the vector sum of the momenta in all calorimeter
cells and $p_T^{\rm max}$ is the transverse momentum of the jet with the
maximum $p_T$. After all these requirements, the
background is reduced to less than 0.1\% in our sample.

The measured energy of each jet formed from calorimeter energy
depositions is not the same as the actual
energy of the particles which enter the calorimeter and
shower.  The jet four-momentum is corrected, on average, to account 
for the energy response of the calorimeters,
the energy showering in and out of the cone, additional
energy from previous beam crossings, and multiple proton-antiproton
interactions in the same event.   The absolute jet energy calibration correction is
determined from the missing transverse energy measured in $\gamma$ + jet
events for the region $|y|\leq 0.4$, while the rapidity dependence is derived from
dijet events using a similar data driven method.
Additionally, since this dijet sample has a large
fraction of gluon initiated jets, corrections of the order of
(2--4)\% are made due to the difference
in response between quark and gluon initiated jets as estimated
using simulated jets produced with the {\sc pythia} event
generator \cite{pythia} that have been passed through a
{\sc geant}-based detector simulation \cite{geant}.
The total jet energy correction varies between 50\% and 20\% for a jet
$p_T$ of 50 to 400~GeV and adjusts the measured
jet energy to the energy of all stable particles that entered the
calorimeter except for muons and neutrinos, which are accounted for
in the final differential cross section.

Bin sizes in $M_{\rm JJ}$ are chosen to be about twice the mass resolution
and to correspond to an efficiency and purity of about 50\% as determined using
a parameterized detector model. The efficiency is defined as
the ratio of Monte Carlo (MC) events 
generated and reconstructed to those generated
in a $M_{\rm{JJ}}$ bin, and purity is defined as the ratio of MC
events generated and reconstructed in a $M_{\rm{JJ}}$ bin
to all events reconstructed in that bin.  The detector model used
is a fast simulation of the D0 detector based on
parameterizations including energy and position resolutions obtained either
from the data or from a detailed simulation of the D0 detector
using {\sc geant}.  This detector model uses events generated by {\sc pythia}
(using the settings of Tune QW~\cite{Albrow:2006rt} 
and MSTW2008LO PDFs~~\cite{MSTW2008}) that have been
reweighted to match measured dijet invariant mass and rapidity distributions
in data. This reweighting assumes a smooth underlying distribution, which does not include resonances. After this tuning, other spectra fundamental to this
measurement, such as the jet $p_T$ distributions, show
good agreement between the data and simulation.
Because the underlying dijet cross sections are steeply falling, the
measured dijet invariant mass distributions are systematically shifted to higher
invariant mass
values due to jet $p_T$ resolution.  The jet $p_T$ resolution
is measured in data using momentum conservation in the transverse
plane for events with exactly two jets,
and is found to be approximately 13\% (7\%) at $p_T \approx 50$ (400) GeV
in the CC and EC, and 16\% (11\%) at $p_T \approx 50$ (400) GeV in the ICR.
The bin-to-bin migrations due to experimental resolution
are determined using the parametrized detector model.
To minimize migrations between $M_{\rm {JJ}}$ bins
due to resolution effects, we use the simulation to obtain a rescaling
function in $M_{\rm{JJ}}$ that optimizes the correlation between the
reconstructed and true values.  The total experimental corrections to the data
are less than $\pm$2\% across the whole dijet invariant mass range
for $|y|_{\rm{max}} < 0.8$, vary
from 0.5\% at $M_{\rm{JJ}} = 0.4$ TeV to 22\% at 1.2 TeV for $0.8 < |y|_{\rm max} < 1.6$,
and from 1\% at $M_{\rm{JJ}} = 0.4$ TeV to 11\% at 1.3 TeV for $1.6 < |y|_{\rm max} < 2.4$.

%\section{\label{sec:result}Results}

%The fully corrected dijet double differential cross section
%is derived from
%\begin{equation}
%d^2\sigma/dM_{\rm JJ}\, d|y|_{\rm max}=(N_{\rm evt}C\,
%(\sum_i (1/\epsilon_{\rm vtx})))/
%({\mathcal L}\, \Delta M_{\rm JJ} \, \Delta|y|_{\rm max})
%\label{eq:mass}
%\end{equation}
%where $N_{\rm evt}$ is the number of events in the mass bin, $\mathcal{L}$ is the
%luminosity, $\epsilon_{\rm vtx}$ is the vertex efficiency per event,
%$\Delta M_{\rm JJ}$ is the mass bin width, $\Delta |y|_{\rm max}$ is
%the rapidity bin width, and $C$ is the correction factor for detector effects
%as described above.

\begin{figure}[htb!]
\includegraphics[width=0.5\textwidth]
{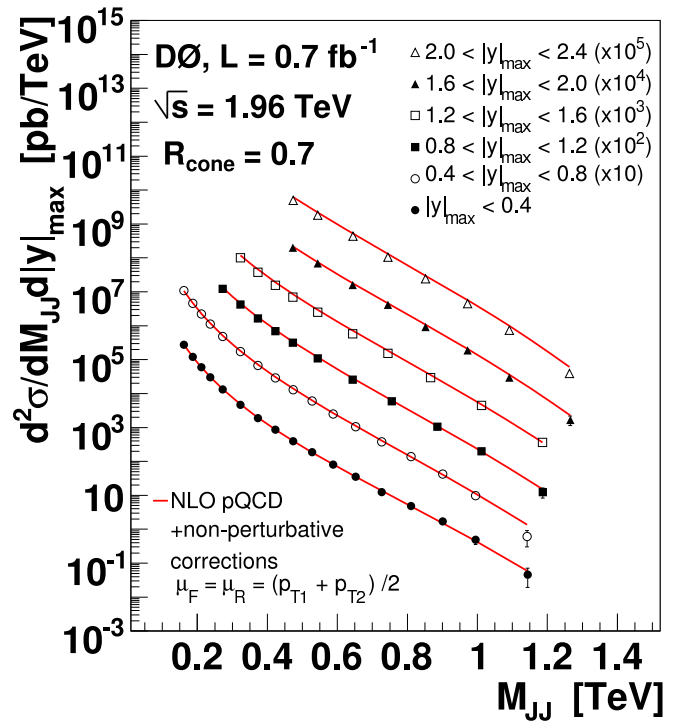}
\caption{(color online) The dijet production cross section as a function of invariant mass
in intervals of $|y|_{\rm max}$ compared to
NLO predictions that include non-perturbative corrections.
The uncertainties shown are statistical only.}
\label{fig:dijet}
\end{figure}

We compute the doubly differential dijet cross section as a function of
dijet invariant mass and $|y|_{\rm max}$ corrected for
all selection efficiencies and migrations due to resolution,
and for the energies of minimum ionizing muons and 
non-interacting neutrinos associated 
with the jet as determined from our detector simulation.  The result is
plotted in all six rapidity regions in
Fig.~\ref{fig:dijet} and tabulated in Tables \ref{tab:cc1nevt}
through \ref{tab:ec2nevt}.
The quoted central value of $M_{\rm JJ}$ in each
bin is the location where the differential cross section
has the same value as the bin average~\cite{LaffertyWyatt:1995}.
%The differential cross section is approximated
%by a parametrization which reproduces the bin averages
%observed in data.

The systematic uncertainties on the cross section are dominated by the uncertainties in the
jet energy calibration,
which range from 6\% to 22\% in the CC, 8\% to 30\% in the ICR, and
15\% to 45\% in the EC region.  The second largest systematic uncertainty
comes from the $p_{T}$ resolution uncertainty, which ranges
between 2\% and 10\% in all regions. The luminosity determination has an uncertainty
of 6.1\%, which is completely correlated across all bins.
The systematic uncertainties on the jet identification
efficiency corrections, corrections due to misvertexing and angular
resolutions, and MC reweighting are
calculated using the parameterized model
of the detector and affect the measured cross
section by less than 2\% in all regions.

\begin{figure*}[htb!]
\centering
\includegraphics[width=1.0\textwidth,angle=0]
{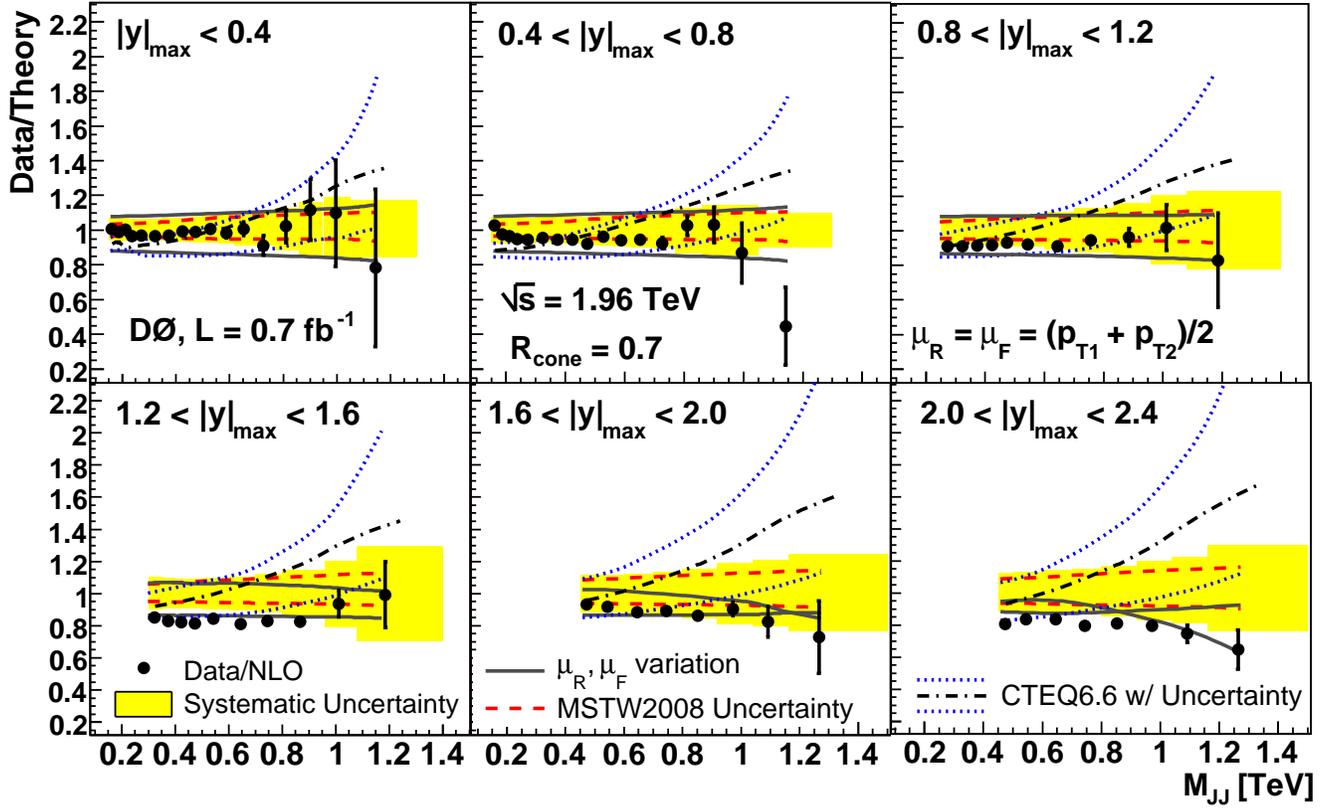}
\caption{(color online) Ratio of data over theoretical expectation
using MSTW2008NLO PDFs in all six $|y|_{\rm max}$ bins. The measurement 
systematic uncertainty is shown as a shaded band.  There is an additional
fully correlated uncertainty of 6.1\% due to the integrated 
luminosity determination which is not shown
in the plots. The legend for all six plots shown is spread out over the three
bottom plots with other relevant information
in the top three plots.  PDF uncertainties show a 90\%~C.L.~band.}
\label{fig:ratio}
\end{figure*}

The data are compared to the next-to-leading
order (NLO) prediction computed using fast{\sc nlo}~\cite{Kluge:2006xs}
based on {\sc nlojet}++ \cite{Nagy:2003tz,Nagy:2001fj}
for MSTW2008NLO PDFs with $\alpha_s(M_Z)=0.120$.
The NLO prediction
is corrected for hadronization and underlying event effects using
corrections which range between
$-10$\% and +23\% depending on the mass in all rapidity regions.
The correction factors are obtained by turning
these effects on and off individually in {\sc pythia}.
The uncertainty due to the nonperturbative corrections
is estimated as 50\% of the individual
corrections, with the uncertainty
determined by adding the individual contributions
in quadrature.
The renormalization and factorization scales are set to
$\mu_{R} = \mu_{F} = p_T = (p_{T1} + p_{T2})/2$ where
$p_{T1}$ and $p_{T2}$  are the $p_T$ of the two highest $p_T$ jets.
The effect of varying these scales simultaneously
from $\mu = p_T/2$ to $\mu = 2p_T$ is shown in Fig.~\ref{fig:ratio}
where the ratio of data to theory is plotted.

\begin{table*}[htb!]
\small
\begin{tabular}{|c c | c c c|c c c c|}
\hline
Mass  & Central & Measured   & Systematic  & Statistical  & Theory  & \multicolumn{3}{c|}{  Non-perturbative  corrections}  \\
range & value &Cross Section  &  uncertainty & uncertainty & Cross Section & Hadron- & Underlying &  Total \\
  TeV    & TeV    &  pb/TeV     &    \%   &   \%     &  pb/TeV  & ization  & event  & \\
  \hline
0.150--0.175 &0.162 &  2.74$\times 10^{5}$ & +7.3,$-$6.6 &  1.9 & 2.74$\times 10^{5}$ &0.917 & 1.180 & 1.082 \\
0.175--0.200 &0.187 &  1.22$\times 10^{5}$ & +7.3,$-$6.6  &  2.6 & 1.22$\times 10^{5}$ &0.930& 1.147 & 1.066\\
0.200--0.225 &0.212 &  6.00$\times 10^{4}$ & +7.3,$-$6.6 & 1.4 & 5.93$\times 10^{4}$ &0.939 & 1.125 & 1.056\\
0.225--0.250 & 0.237 &  3.02$\times 10^{4}$ & +7.3,$-$6.6 & 1.8 & 3.10$\times 10^{4}$ &0.945 & 1.110  & 1.049 \\
0.250--0.300 & 0.272 &  1.32$\times 10^{4}$ & +7.3,$-$6.6 &  1.3 & 1.36$\times 10^{4}$ &0.950 & 1.095  & 1.041\\
0.300--0.350 & 0.323 &  4.69$\times 10^{3}$ & +7.5,$-$6.8 & 1.6 & 4.85$\times 10^{3}$ &0.955 & 1.083 & 1.035 \\
0.350--0.400 & 0.373 &  1.90$\times 10^{3}$ & +7.3,$-$6.7 & 1.3 & 1.96$\times 10^{3}$ &0.959 &  1.075 &1.030 \\
0.400--0.450 & 0.423 &  8.48$\times 10^{2}$ & +7.4,$-$6.8 & 1.4 & 8.60$\times 10^{2}$ &0.961 & 1.069 &  1.027 \\
0.450--0.500 & 0.473 &  3.93$\times 10^{2}$ & +7.6,$-$7.1  & 1.7 & 4.01$\times 10^{2}$ &0.963 & 1.065 & 1.025 \\
0.500--0.560 & 0.528 &  1.84$\times 10^{2}$ & +7.9,$-$7.4 &  2.1 & 1.85$\times 10^{2}$ &0.965 & 1.058 & 1.022 \\
0.560--0.620 &0.588 &  7.93$\times 10^{1}$ & +8.3,$-$8.0 &3.1 & 8.17$\times 10^{1}$ &0.967 & 1.054 &  1.019 \\
0.620--0.690& 0.652 &  3.50$\times 10^{1}$ & +9.1,$-$8.8 &  4.2 & 3.53$\times 10^{1}$ &0.966 & 1.056 &1.020 \\
0.690--0.770 & 0.727 &  1.23$\times 10^{1}$ & +10.4,$-$10.0 &  6.5 & 1.37$\times 10^{1}$ &0.967 & 1.054  & 1.019 \\
0.770--0.860 & 0.811 &  4.83$\times 10^{0}$ & +12.1,$-$11.7 & 9.8 & 4.77$\times 10^{0}$ &0.968 &  1.052  & 1.018 \\
0.860--0.950 & 0.901 &  1.69$\times 10^{0}$ & +14.3,$-$13.7 & 15.8 & 1.52$\times 10^{0}$ & 0.968 & 1.050 & 1.017 \\
0.950--1.050& 0.995 &  4.95$\times 10^{-1}$ & +16.7,$-$15.8 & 31.6 & 4.49$\times 10^{-1}$ &0.969 & 1.049  & 1.016 \\
1.050--1.300 & 1.144 &  4.56$\times 10^{-2}$ & +22.1,$-$20.0 & 57.7 & 5.83$\times 10^{-2}$ &0.970 &  1.047 &1.015 \\

\hline
\end{tabular}
\caption[Dijet double differential cross section, $d^2\sigma/dMd|y|_{\rm max}$, for $|y|_{\rm max} \leq 0.4$]{\label{tab:cc1nevt} Dijet double differential cross section, $d^2\sigma/dMd|y|_{\rm max}$, for $|y|_{\rm max} \leq 0.4$, compared to theoretical predictions with non-perturbative corrections.  There is an additional
fully correlated uncertainty of 6.1\% due to the integrated luminosity determination which is not shown
in the table.}
\end{table*}

\begin{table*}[htb!]
\small
\begin{tabular}{|c c | c c c| c c c c|}
\hline
Mass  & Central & Measured   & Systematic  & Statistical  & Theory  & \multicolumn{3}{c|}{  Non-perturbative  corrections}  \\
range & value &Cross Section  &  uncertainty & uncertainty & Cross Section & Hadron- & Underlying &  Total \\
  TeV    & TeV    &  pb/TeV     &    \%   &   \%     &  pb/TeV  & ization  & event  & \\
  \hline
0.150--0.175 & 0.162 &  1.08$\times 10^{6}$ & +7.4,$-$7.4 & 1.3 & 1.07$\times 10^{6}$ &0.946 &1.127 &  1.066  \\
0.175--0.200 &0.187 &  4.67$\times 10^{5}$ & +7.5,$-$7.4& 1.6 & 4.73$\times 10^{5}$ &0.951 & 1.109 &  1.055 \\
0.200--0.225 & 0.212 &  2.24$\times 10^{5}$ & +7.5,$-$7.5 &  1.1 & 2.29$\times 10^{5}$ &0.955 &1.094 &  1.045  \\
0.225--0.250 & 0.237 &  1.14$\times 10^{5}$ & +7.6,$-$7.5 &  1.2 & 1.19$\times 10^{5}$ &0.958 & 1.084 & 1.040 \\
0.250--0.300 & 0.272 &  4.91$\times 10^{4}$ & +7.9,$-$7.8 & 1.1 & 5.14$\times 10^{4}$ &0.960 &  1.077 & 1.034\\
0.300--0.350 &0.323 &  1.74$\times 10^{4}$ & +7.6,$-$7.6 &  1.2 & 1.81$\times 10^{4}$ &0.961 &  1.072 &1.030 \\
0.350--0.400 &0.373 &  6.77$\times 10^{3}$ & +7.9,$-$7.7 & 1.1 & 7.15$\times 10^{3}$ &0.963 & 1.067& 1.028 \\
0.400--0.450 &0.423 &  2.89$\times 10^{3}$ & +8.0,$-$7.9 &  1.2 & 3.07$\times 10^{3}$ &0.964 & 1.064 &1.025\\
0.450--0.500 &0.473 &  1.28$\times 10^{3}$ & +8.3,$-$8.2 &  1.3 & 1.40$\times 10^{3}$ &0.964 & 1.061 &1.023 \\
0.500--0.560 & 0.528 &  5.97$\times 10^{2}$ & +8.7,$-$8.6 & 1.4 & 6.25$\times 10^{2}$ &0.965 & 1.058 & 1.021 \\
0.560--0.620 &0.589 &  2.50$\times 10^{2}$ & +9.4,$-$9.2 &  1.9 & 2.68$\times 10^{2}$ &0.966 & 1.056 & 1.020 \\
0.620--0.690 &0.652 &  1.04$\times 10^{2}$ & +10.3,$-$10.1 & 2.5 & 1.11$\times 10^{2}$ &0.966 &  1.054 &1.018  \\
0.690--0.770 &0.726 &  3.78$\times 10^{1}$ & +11.7,$-$11.3 &  3.8 & 4.12$\times 10^{1}$ &0.967 & 1.052  & 1.017 \\
0.770--0.860 & 0.811 &  1.38$\times 10^{1}$ & +13.5,$-$13.0 &  5.7 & 1.35$\times 10^{1}$ &0.967& 1.050& 1.016 \\
0.860--0.950 &0.901 &  4.20$\times 10^{0}$ & +15.7$,-$14.9 &  10.7 & 4.08$\times 10^{0}$ &0.968 & 1.047 & 1.014 \\
0.950--1.050 &0.994 &  9.90$\times 10^{-1}$ & +18.4,$-$17.0 &  20.4 & 1.13$\times 10^{0}$ &0.969 & 1.045 & 1.012\\
1.050--1.300 &1.142 &  6.08$\times 10^{-2}$ & +23.5,$-$20.9 &  50.0 & 1.36$\times 10^{-1}$ & 0.969 &  1.045 &1.012 \\
\hline
\end{tabular}
\caption[Dijet invariant mass cross section for $ 0.4 < |y|_{\rm max} \leq 0.8$]{\label{tab:cc2nevt}  Dijet double differential cross section, $d^2\sigma/dMd|y|_{\rm max}$, for $ 0.4 < |y|_{\rm max} < 0.8$, compared to theoretical predictions with
non-perturbative corrections. There is an additional
fully correlated uncertainty of 6.1\% due to the integrated luminosity determination which is not shown
in the table.}
\end{table*}

\begin{table*}[htb!]
\small
\begin{tabular}{|c c | c c c| c c c c|}
\hline
Mass  & Central & Measured   & Systematic  & Statistical  & Theory  & \multicolumn{3}{c|}{  Non-perturbative  corrections}  \\
range & value &Cross Section  &  uncertainty & uncertainty & Cross Section & Hadron- & Underlying &  Total \\
  TeV    & TeV    &  pb/TeV     &    \%   &   \%     &  pb/TeV  & ization  & event  & \\
  \hline
0.250--0.300 &0.272 &  1.21$\times 10^{5}$ & +10.3,$-$10.0 &  1.1& 1.34$\times 10^{5}$ &0.949 & 1.126 & 1.069 \\
0.300--0.350 &0.323 &  4.18$\times 10^{4}$ & +9.7,$-$9.5 &  1.3 & 4.63$\times 10^{4}$ &0.953 & 1.111 & 1.059  \\
0.350--0.400 & 0.373 &  1.63$\times 10^{4}$ & +9.4,$-$9.1 &  1.7 & 1.80$\times 10^{4}$ &0.956 & 1.100 & 1.052 \\
0.400--0.450 &0.423 &  6.86$\times 10^{3}$ & +9.3,$-$9.0  &  1.4 & 7.55$\times 10^{3}$ &0.958 & 1.092 & 1.046 \\
0.450-- 0.500 &0.473 &  3.10$\times 10^{3}$ & +9.3,$-$9.0 &  1.9 & 3.38$\times 10^{3}$ &0.960 & 1.083 & 1.041 \\
0.500--0.600 &0.544 &  1.07$\times 10^{3}$ & +9.6,$-$9.3 & 1.2 & 1.17$\times 10^{3}$ &0.963 & 1.076 &  1.035  \\
0.600--0.700 &0.644 &  2.57$\times 10^{2}$ & +10.6,$-$10.4 &  1.8 & 2.83$\times 10^{2}$ &0.964 & 1.070 &1.031 \\
0.700--0.830 &0.756 &  5.95$\times 10^{1}$ & +12.7,$-$12.6 &  2.5 & 6.30$\times 10^{1}$ &0.965 &  1.065 &1.028 \\
0.830--0.960 & 0.886 &  1.08$\times 10^{1}$ & +16.4,$-$16.0& 5.4 & 1.10$\times 10^{1}$ &0.966 & 1.062 & 1.026 \\
0.960--1.080 &1.012 &  2.10$\times 10^{0}$ & +20.6,$-$19.7 &  12.5 & 1.95$\times 10^{0}$ &0.967 & 1.058  &  1.023 \\
1.080--1.400 &1.186 &  1.43$\times 10^{-1}$ & +28.5,$-$24.5 & 28.9 & 1.50$\times 10^{-1}$ &0.969 &  1.053  & 1.020  \\
 \hline

\end{tabular}
\caption[Dijet invariant mass cross section for $ 0.8 < |y|_{\rm max} \leq 1.2$]{\label{tab:ic1nevt} Dijet double differential cross section, $d^2\sigma/dMd|y|_{\rm max}$, for $ 0.8 < |y|_{\rm max} \leq1.2$, compared to theoretical predictions with
non-perturbative corrections. There is an additional
fully correlated uncertainty of 6.1\% due to the integrated luminosity determination which is not shown
in the table.}
\end{table*}

\begin{table*}[htb!]
\small
\begin{tabular}{|c c | c c c| c c c c|}
\hline
Mass  & Central & Measured   & Systematic  & Statistical  & Theory  & \multicolumn{3}{c|}{  Non-perturbative  corrections}  \\
range & value &Cross Section  &  uncertainty & uncertainty & Cross Section & Hadron- & Underlying &  Total \\
  TeV    & TeV    &  pb/TeV     &    \%   &   \%     &  pb/TeV  & ization  & event  & \\
  \hline
0.300--0.350 &0.323 &  1.00$\times 10^{5}$ & +10.7,$-$10.4 & 1.2 & 1.19$\times 10^{5}$ &0.949 &  1.143 &1.085 \\
0.350--0.400 &0.373 &  3.79$\times 10^{4}$ & +10.4,$-$10.1 &  1.3 & 4.60$\times 10^{4}$ &0.951 &  1.133 &  1.077 \\
0.400--0.450 & 0.423 &  1.61$\times 10^{4}$ & +10.4,$-$9.9 &  1.7 & 1.91$\times 10^{4}$ &0.952 & 1.125 &  1.071 \\
0.450--0.500 & 0.473 &  7.11$\times 10^{3}$ & +10.7,$-$10.0  &  2.3 & 8.60$\times 10^{3}$ &0.954 & 1.116  &1.065 \\
0.500--0.600 &0.544 &  2.54$\times 10^{3}$ & +11.3,$-$10.4 & 1.6 & 2.97$\times 10^{3}$ &0.955 &  1.109  & 1.059  \\
0.600--0.700 &0.644 &  5.94$\times 10^{2}$ & +12.3,$-$11.7 &  1.3 & 7.16$\times 10^{2}$ &0.956 & 1.103 &  1.055 \\
0.700--0.800 &0.744 &  1.58$\times 10^{2}$ & +14.1,$-$13.4 &  2.1 & 1.84$\times 10^{2}$ &0.957 & 1.098 & 1.051 \\
0.800--0.960 &0.866 &  3.16$\times 10^{1}$ & +17.8,$-$16.8 &  2.9 & 3.57$\times 10^{1}$ &0.958 & 1.095& 1.048 \\
0.960--1.080 & 1.012 &  5.08$\times 10^{0}$ & +22.7,$-$21.4 &  8.0 & 4.78$\times 10^{0}$ &0.958 &  1.091  & 1.045  \\
1.080--1.400 & 1.186 &  4.77$\times 10^{-1}$ & +29.5,$-$27.9& 15.8 & 3.67$\times 10^{-1}$ &0.959 & 1.084   &1.040 \\
 \hline
\end{tabular}
\caption[Dijet invariant mass cross section for $ 1.2 < |y|_{\rm max} \leq 1.6$]{\label{tab:ic2nevt} Dijet double differential cross section, $d^2\sigma/dMd|y|_{\rm max}$, for $ 1.2 < |y|_{\rm max} \leq 1.6$, compared to theoretical predictions with
non-perturbative corrections. There is an additional
fully correlated uncertainty of 6.1\% due to the integrated luminosity determination which is not shown
in the table.}
\end{table*}

\begin{table*}[htb!]
\small
\begin{tabular}{|c c | c c c| c c c c|}
\hline
Mass  & Central & Measured   & Systematic  & Statistical  & Theory  & \multicolumn{3}{c|}{  Non-perturbative  corrections}  \\
range & value &Cross Section  &  uncertainty & uncertainty & Cross Section & Hadron- & Underlying &  Total \\
  TeV    & TeV    &  pb/TeV     &    \%   &   \%     &  pb/TeV  & ization  & event  & \\
  \hline
0.450--0.500 &0.473 &  2.01$\times 10^{4}$ & +12.0$-$13.5 & 2.2 & 2.27$\times 10^{4}$ &0.940 &1.151 & 1.083 \\
0.500--0.600 &0.544 &  6.88$\times 10^{3}$ & +13.8,$-$14.6 &  2.3 & 7.82$\times 10^{3}$ &0.940 & 1.141 &  1.073\\
0.600--0.700 &0.644 &  1.58$\times 10^{3}$ & +16.3,$-$17.3 & 3.2 & 1.87$\times 10^{3}$ &0.941 & 1.132 & 1.065  \\
0.700--0.800 &0.744 &  4.10$\times 10^{2}$ & +19.9,$-$18.7 &  2.3 & 4.74$\times 10^{2}$ &0.941 &1.125 & 1.058 \\
0.800--0.920 &0.852 &  9.30$\times 10^{1}$ & +21.1,$-$17.0 & 2.8 & 1.10$\times 10^{2}$ &0.941 &  1.119  &1.054 \\
0.920--1.040 &0.972 &  1.93$\times 10^{1}$ & +27.1,$-$20.3 & 4.9 & 2.16$\times 10^{1}$ &0.941 &  1.112 &1.047 \\
1.040--1.160 &1.092 &  3.15$\times 10^{0}$ & +32.5,$-$24.3  & 11.2 & 3.68$\times 10^{0}$ &0.942 & 1.104 &  1.040\\
1.160--1.500 & 1.266 &  1.92$\times 10^{-1}$ & +36.3,$-$33.4 &  25.1 & 2.34$\times 10^{-1}$ &0.942 & 1.100 & 1.037\\
 \hline
\end{tabular}
\caption[Dijet invariant mass cross section for $ 1.6 < |y|_{\rm max} \leq 2.0$]{\label{tab:ec1nevt} Dijet double differential cross section, $d^2\sigma/dMd|y|_{\rm max}$, for $ 1.6 < |y|_{\rm max} \leq 2.0$, compared to theoretical predictions with
non-perturbative corrections. There is an additional
fully correlated uncertainty of 6.1\% due to the integrated luminosity determination which is not shown
in the table.}
\end{table*}

\begin{table*}[htb!]
\small
\begin{tabular}{|c c | c c c| c c c c|}
\hline
Mass  & Central & Measured   & Systematic  & Statistical  & Theory  & \multicolumn{3}{c|}{  Non-perturbative  corrections}  \\
range & value &Cross Section  &  uncertainty & uncertainty & Cross Section & Hadron- & Underlying &  Total \\
  TeV    & TeV    &  pb/TeV     &    \%   &   \%     &  pb/TeV  & ization  & event  & \\
  \hline
0.450--0.500 &0.473 &  4.95$\times 10^{4}$ & +16.1,$-$13.7 &  2.1 & 6.08$\times 10^{4}$ &0.928 & 1.229& 1.141 \\
0.500--0.600 &0.544 &  1.81$\times 10^{4}$ & +16.2,$-$14.1 &  2.1 & 2.15$\times 10^{4}$ &0.925 &  1.222 & 1.130\\
0.600--0.700 &0.644 &  4.36$\times 10^{3}$ & +16.5,$-$15.2 &  2.5 & 5.21$\times 10^{3}$ &0.923 &  1.216 &1.122\\
0.700--0.800 &0.744 &  1.02$\times 10^{3}$ & +17.4,$-$17.0 &  2.1 & 1.31$\times 10^{3}$ &0.920 &  1.211   & 1.115  \\
0.800--0.920 & 0.852 &  2.37$\times 10^{2}$ & +20.0,$-$19.9 &  2.4 & 2.998$\times 10^{2}$ &0.919 & 1.208   &  1.110  \\
0.920--1.040 & 0.972 &  4.43$\times 10^{1}$ & +24.8,$-$23.9 &  3.5 & 5.66$\times 10^{1}$ &0.917 &1.203  & 1.103 \\
1.040--1.160 &1.091 &  7.25$\times 10^{0}$ & +33.0,$-$28.0 &  7.3 & 9.86$\times 10^{0}$ &0.915 & 1.198&  1.095\\
1.160--1.500 &1.263 &  4.12$\times 10^{-1}$ & +46.1,$-$33.8 &  16.5 & 6.09$\times 10^{-1}$ &0.913 & 1.195 & 1.092 \\
 \hline
\end{tabular}
\caption[Dijet invariant mass cross section for $ 2.0 < |y|_{\rm max} \leq 2.4$]{\label{tab:ec2nevt} Dijet double differential cross section, $d^2\sigma/dMd|y|_{\rm max}$, for $ 2.0 < |y|_{\rm max} \leq 2.4$, compared to theoretical predictions with
non-perturbative corrections. There is an additional
fully correlated uncertainty of 6.1\% due to the integrated luminosity determination which is not shown
in the table.}
\end{table*}

The experimental uncertainties are similar in size to both the 
PDF and the scale
uncertainties, suggesting that the measurement will constrain 
theoretical models.  
We are quoting PDF uncertainties corresponding to a 90\% C.L.
The total uncertainties are smaller than
those of earlier measurements at this same center-of-mass energy \cite{cdfrun2}.
In addition to comparing the D0 measurement to the
theoretical predictions using MSTW2008NLO PDFs,
we also compare to the
theoretical predictions using CTEQ6.6 PDFs \cite{cteq66}.
The difference in the cross section due to the choice of PDFs
is  (40--60)\% at the highest mass.  Although the central value for
the MSTW2008NLO PDFs are favored, it is important to note that their
determination included a measurement of the D0 inclusive jet production
cross section \cite{d0jet} which is based on the same dataset as the 
present measurement. In addition, these PDFs exclude Tevatron data taken 
before 2000, while the CTEQ6.6 PDFs include that data and do not 
include Tevatron data taken after 2000.

In summary, we have presented a new measurement of the dijet production
cross section as a function of the dijet invariant mass and of the 
largest rapidity of the two highest $p_T$ jets
that extends the rapidity range beyond previous measurements, with
systematic uncertainties that are significantly smaller.
In general, the data are described by NLO QCD predictions
using MSTW2008NLO or CTEQ6.6 PDFs in all rapidity regions,  
though the central value of the CTEQ6.6 PDFs differs 
from the data for high dijet mass at larger rapidities.

\input acknowledgement_paragraph_r2.tex   % input acknowledgement

\end{document}

%% file: list_of_authors_r2.tex
% LIST_OF_AUTHORS_R2.TEX      4 Feb 2010                    
%
\author{V.M.~Abazov$^{36}$}
\author{B.~Abbott$^{74}$}
\author{M.~Abolins$^{63}$}
\author{B.S.~Acharya$^{29}$}
\author{M.~Adams$^{49}$}
\author{T.~Adams$^{47}$}
\author{E.~Aguilo$^{6}$}
\author{G.D.~Alexeev$^{36}$}
\author{G.~Alkhazov$^{40}$}
\author{A.~Alton$^{62,a}$}
% ...

\affiliation{\vspace{0.1 in}(The D\O\ Collaboration)\vspace{0.1 in}}
% ...
\affiliation{$^{6}$Simon Fraser University, Burnaby, British Columbia, Canada;
                and York University, Toronto, Ontario, Canada}
% ...
\affiliation{$^{29}$Tata Institute of Fundamental Research, Mumbai, India}
% ...
\affiliation{$^{36}$Joint Institute for Nuclear Research, Dubna, Russia}
% ...
\affiliation{$^{40}$Petersburg Nuclear Physics Institute,
                St. Petersburg, Russia}
% ...
\affiliation{$^{47}$Florida State University, Tallahassee, Florida 32306, USA}
% ...
\affiliation{$^{49}$University of Illinois at Chicago,
                Chicago, Illinois 60607, USA}
% ...
\affiliation{$^{62}$University of Michigan, Ann Arbor, Michigan 48109, USA}
\affiliation{$^{63}$Michigan State University,
                East Lansing, Michigan 48824, USA}
% ...
\affiliation{$^{74}$University of Oklahoma, Norman, Oklahoma 73019, USA}
% ...

%% file: acknowledgement_paragraph_r2.tex
% acknowledgement_paragraph_r2.tex                          2/4/10
%
We thank the staffs at Fermilab and collaborating institutions, 
and acknowledge support from the 
DOE and NSF (USA);
CEA and CNRS/IN2P3 (France);
FASI, Rosatom and RFBR (Russia);
CNPq, FAPERJ, FAPESP and FUNDUNESP (Brazil);
DAE and DST (India);
Colciencias (Colombia);
CONACyT (Mexico);
KRF and KOSEF (Korea);
CONICET and UBACyT (Argentina);
FOM (The Netherlands);
STFC and the Royal Society (United Kingdom);
MSMT and GACR (Czech Republic);
CRC Program and NSERC (Canada);
BMBF and DFG (Germany);
SFI (Ireland);
The Swedish Research Council (Sweden);
and
CAS and CNSF (China).